# An Empirical Investigation of V-I Trajectory based Load Signatures for Non-Intrusive Load Monitoring

T. Hassan, F. Javed and N. Arshad, *Member, IEEE*

*Abstract*-- Choice of load signature or feature space is one of the most fundamental design choices for non-intrusive load monitoring or energy disaggregation problem. Electrical power quantities, harmonic load characteristics, canonical transient and steady-state waveforms are some of the typical choices of load signature or load signature basis for current research addressing appliance classification and prediction. This paper extends and evaluates appliance load signatures based on V-I trajectory – the mutual locus of instantaneous voltage and current waveforms – for precision and robustness of prediction in classification algorithms used to disaggregate residential overall energy use and predict constituent appliance profiles. We also demonstrate the use of variants of differential evolution as a novel strategy for selection of optimal load models in context of energy disaggregation. A publically available benchmark dataset REDD is employed for evaluation purposes. Our experimental evaluations indicate that these load signatures, in conjunction with a number of popular classification algorithms, offer better or generally comparable overall precision of prediction, robustness and reliability against dynamic, noisy and highly similar load signatures with reference to electrical power quantities and harmonic content. Herein, wave-shape features are found to be an effective new basis of classification and prediction for semi-automated energy disaggregation and monitoring.

## I. INTRODUCTION

Non-Intrusive Load Monitoring (NILM) is the disaggregation of overall demand profile of a household into individual signatures of appliances switched on at a particular instant or within a specified time period. This disaggregation is carried out without using intrusive physical sensors on individual appliances. Aggregate demand profile (ADP) or instantaneous power consumption is typically observed via a single energy meter installed at service entry point of the household. Load signatures of individual appliances refer to metrics that characterize their operating state and temporal behavior. This study evaluates a new kind of two-dimensional load signature for non-intrusive profiling and identification of residential appliances. Load signatures are essential to profiling, status monitoring and safety assurance of electrical loads ([3], [5]). Our design choice for load signatures is wave-shape features (WS) based on the mutual trajectory of instantaneous voltage and current waveforms. Lam and colleagues originally introduced these load signatures as a new basis for establishing and comparing load taxonomies [3]. In this study, in contrast, we have evaluated WS for precision and robustness of prediction in NILM as a multi-class classification problem. Our empirical results illustrate that precision of prediction and robustness against dynamic, noisy and highly similar load signatures in NILM with wave-shape features (WS) is better or generally comparable to that of traditional benchmark load signatures over a variety of classification algorithms. To evaluate our algorithms we have used a publically available dataset for benchmarking i.e. Reference Energy Disaggregation Dataset (REDD) [14].

## II. RELATED WORK

Contemporary research on implementation of a NILM system typically addresses the following design choices.

**Granularity over time of ADP**: Granularity over ADP refers to the rate at which the installed meter is able to observe and report the overall instantaneous power consumption. This design choice is usually associated with the distinction between event-based and non-event based operation. In event-based operation, decision frequency for NILM is the frequency at which appliances in the household change their state ([8], [29]-[30]). This choice is appropriate for real-time diagnostic feedback in residential and commercial energy feedback systems (REFS and CEFS) and is considered for this paper. Non-event based NILM operates on aggregated consumption data and is useful for long-term or medium term diagnostics on energy consumption of particular households or for larger settings such as microgrids and distribution sectors, particularly as an enabling technology for demand-side management. [34]

**Definitions of load signatures:** As introduced earlier, the definitions of load signatures refer to the metrics used to identify operating characteristics of individual devices connected within the setting in question. Various existing implementations of NILM systems primarily differ in choice of steady-state or transient, fundamental frequency or harmonic frequency signatures. For instance, Hart [1], Cole and Albicki [7] employ steady-state and transient changes of

This work is supported by National ICT R&D Fund, Ministry of Information Technology, Government of Pakistan.

T. Hassan, F. Javed and N. Arshad are with the Department of Computer Science at Lahore University of Management Sciences, Lahore, Pakistan. E-mail: {taha.hassan, fahadjaved, naveedarshad}@lums.edu.pk.

active real power as basis of classification. In contrast, Lee et. al. [30] and Laughman et. al. [29] employ harmonic magnitudes associated with step changes in overall load for establishing and comparing appliance profiles. Liang and colleagues [8] have demonstrated the use of raw single-cycle current, instantaneous power and admittance waveforms for the purpose. A comprehensive review of signature types evaluated for NILM appears in surveys by Zeifman [28] and Ahmad et. al. [27]. Lam et. al. [3] study typical bases for load taxonomies and introduce wave-shape features (WS) as a competitive new basis using hierarchical clustering. According to the study, WS characterize the shape of instantaneous current demand for a particular device, carry engineering meanings and result in larger relative differentiation between appliances with different operating principles. A prior examination of WS for NILM however, does not exist to best of the authors' knowledge and hence the subject of this paper.

**Initial acquisition of load signatures**: Hart's canonical work refers to NILM's two distinct modes of operation, with different degrees of intrusiveness [1]. The first uses a one-time calibration period where appliance signatures are manually collected for supervised learning algorithms. The mode is referred to as 'Manual-Setup' (MS-NILM) or semi-automated energy disaggregation and is considered for this paper. Hart [1], Cole and Albicki [7] and Liang et. al. [8] are typical implementations of MS-NILM. The second operational mode for NILM, termed 'Automatic-Setup', uses *a priori* information about expected load characteristics and unsupervised machine learning algorithms to automatically disaggregate ADP. Contributions by Parson et. al. [25] and Kim [26] are recent investigations of unsupervised energy disaggregation. MS-NILM is computationally convenient, however, the intrusive collection and labeling of signatures makes it tedious to set up and adapt to new appliances.

**Selection of learning and optimization algorithms:** This refers to the parameter search and performance optimization of learning algorithms, for instance, artificial neural network or support vector method employed to learn appliance signatures. This choice of algorithms for disaggregation is also fundamentally relative to both the required response time for NILM system and the choice of load signatures. A wide variety of supervised and unsupervised learning algorithms ([7]-[8], [16]-[17], [25]-[26]) and optimization strategies such as integer programming and metaheuristics ([9]-[12]) have been observed effective in their capacity for load profiling and disaggregation. This capacity is typically distinct relative to load conditions at the time of operation, for instance, number of simultaneously operating appliances, noise levels, electrical interference from neighboring system, etc. [8].

Our contributions in this paper can be summarized as follows.
1) We expand the set of load characteristics in WS and present a comprehensive empirical evaluation of WS as basis for profiling, and prediction of appliances in manual-setup, event-based non-intrusive load monitoring. We identify and compare (a) the capacity of load disaggregation and (b) its robustness against dynamic and highly similar load signatures. Our experiments also take into account ambient variation in load signatures due to electrical noise and interference typical of household ADP.
2) Additionally, we select optimal values of adjustable parameters for employed learning algorithms using population-based global search. The algorithm is based on differential evolution (DE) [6], which is a novel technique in the context of NILM.
3) Finally, we present an overview of disaggregation capacity of employed learning algorithms as a function of aforementioned conditions, utilizing reference energy disaggregation dataset (REDD) [14] as benchmark dataset for evaluation.

III. DESCRIPTION OF DESIGN METHODOLOGY

A system-level depiction of typical manual-setup NILM is presented in figure 1. 'Manual-setup' refers to a one-time calibration period to learn the appliances signatures and store them in a database. Once the system learns these signatures it is able to identify the appliances in the system whenever a switching event takes place.

There are usually two forms of signatures used in NILM i.e. snapshot-form signatures and delta-form signatures [2]. Snapshot form refers to signatures that are the aggregate power consumption of all appliances as observed via energy meter installed for NILM. On the other hand delta-form signatures express load behavior in brief windows of time containing only a single switching event rather than a large number of events to facilitate event-based NILM. For the scope of this paper only delta-form signatures are considered.

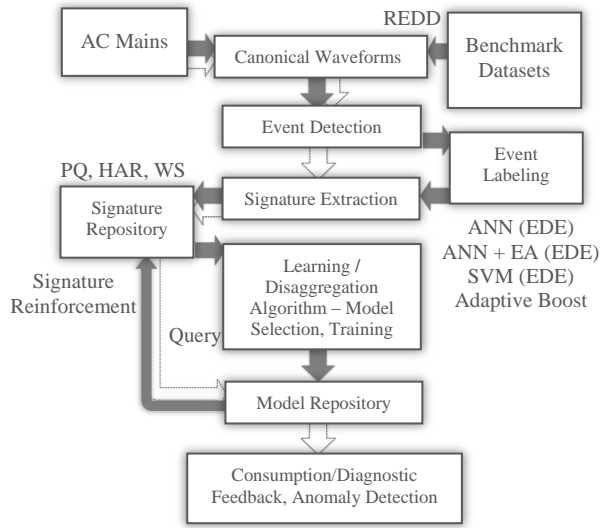

Fig. 1. System-level depiction of manual-setup non-intrusive load disaggregation. Dashed trajectory refers to NILM query; solid refers to NILM training and optimization.

*A. Pre-processing Energy Consumption Data*

For our evaluation, we need energy consumption data for the setting where NILM is deployed. For high frequency approaches to NILM such as this study, acquisition of instantaneous voltage and current waveforms is set up, typically at a rate of more than 100 samples per cycle. This data is used to make the system learn the load signatures of various appliances. Wave-shape features evaluated in the following sections use the mutual trajectory of voltage and current waveforms to deduce information about electrical appliances, hence the high granularity. For the course of this study, benchmark dataset REDD [14] is used as a source of these raw waveforms. Detail on denominations in REDD is mentioned later in the study.

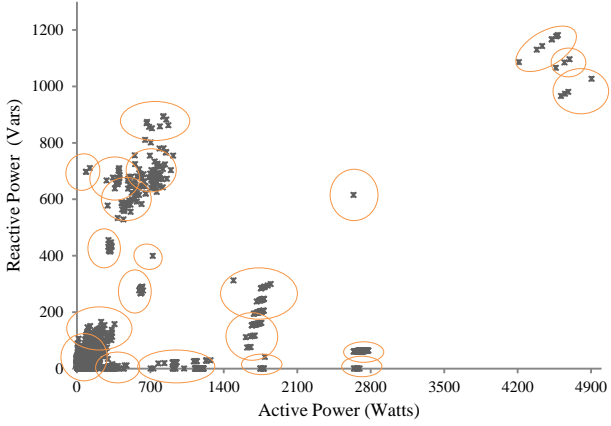

Fig. 2. Consumption values from a subset of switching events from REDD – traditional power metrics ($P, Q$).

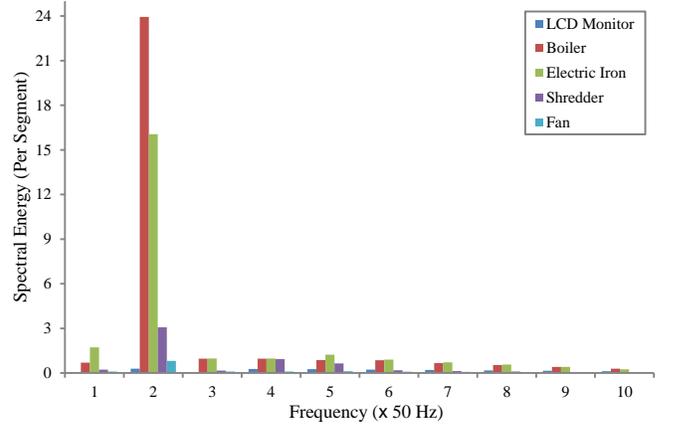

Fig. 3. Consumption values for various appliances - harmonic content (HAR).

Pre-processing of these raw waveforms includes separation of switching events, which are time instances at which a particular appliance in the household changes state. Once those instances are determined, cycle-by-cycle snapshots of both voltage and current waveforms are extracted, otherwise known as delta-form signatures [2]. These switching events are 'undifferentiated' in that these delta-form signatures have not been attributed to individual appliances that they represent so far. In order to evaluate WS for NILM as multi-class classification, K-means clustering is used to group these signatures into cohesive groups with unique appliance IDs.

### B. Standard Benchmark Load Signatures

The first benchmark used in our evaluation is traditional power metrics (PQ) [1]. PQ refers to real and reactive power consumption of a particular appliance ($P, Q$) and total odd and even harmonic distortion of current waveform ($THD_o, THD_e$).

The second benchmark used is harmonic content (HAR) of the current waveforms, determined by the spectral energy in contiguous segments of Fourier Transform (FT) of the current waveform. Fig. 2 represents the power consumption of various appliances from one particular household in our example dataset REDD on a $P - Q$ plane for 5000 consecutive switching events. Note the abundance of low-power (<200W) appliance signatures in close proximity. Fig. 3 illustrates normalized coefficients that represent the energy in various frequency bands of the current waveform spectrum for five common appliances.

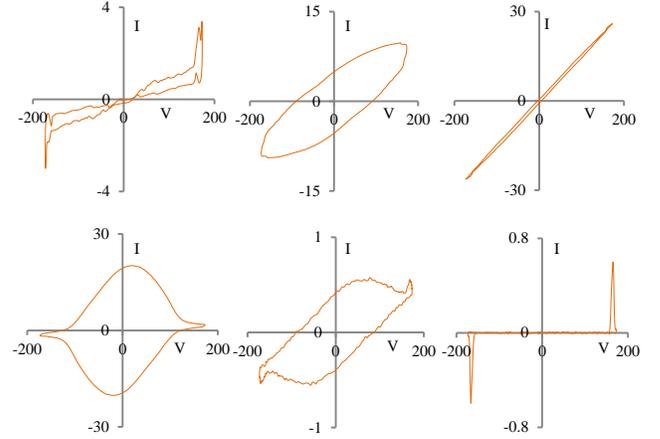

Fig. 4. A graphical illustration of V-I trajectories for six different appliances from REDD household # 3.

### C. Wave-shape Features (WS)

Lam and colleagues introduced metrics based on voltage and current wave-shape for establishing taxonomy of appliance signatures [3]. Lam's work refers to the mutual locus of instantaneous voltage and current waveform as the V-I trajectory. Fig. 4 illustrates the shape of V-I trajectory for six different appliances from one of REDD households. For appliances with different working principles (for instance, resistive, motor-driven or power-electronic), the V-I trajectory exhibits different unique characteristics that are captured by wave-shape (WS) features. Since operating state of a particular appliance corresponds to shape of instantaneous current waveform of that particular appliance, hierarchical clustering with WS results in larger separation between characteristically dissimilar appliances compared to PQ. An extrapolation of this observation is to verify that WS would allow a multi-class classifier to generalize better to unknown examples. This would prove the effectiveness of WS for load disaggregation over a variety of off-the-shelf learning algorithms and benchmark load signatures.

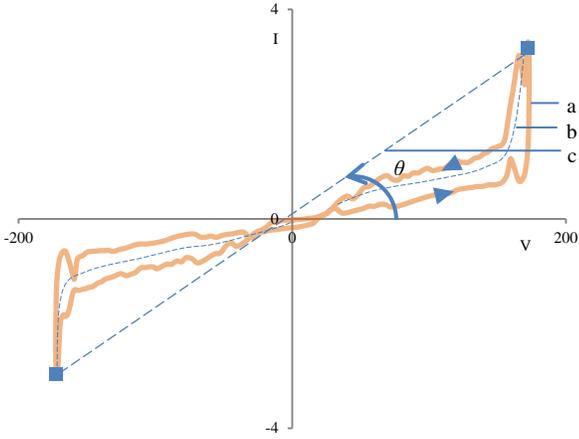

Fig. 5. A graphical illustration of wave-shape metrics: (*a*) V-I trajectory (*b*) Mean curve (*c*) Reference line joining points of highest and lowest I-coordinate in the V-I plane.

A subset of proposed WS used for this study is briefly reviewed as follows; detailed explanation appears in [3].

**Looping Direction:** 'Looping direction' refers to the anti-clockwise or clockwise curvature of V-I trajectory. It corresponds to the sign of the phase angle difference between voltage and current waveforms, represented in fig. 5 by the direction of curvature of locus *a*. A clockwise curvature refers to an overall capacitive load behavior (current leads voltage in phase) while a counter-clockwise curvature refers to an overall inductive load characteristic (voltage leads current in phase).

**Area Enclosed:** 'Area Enclosed' refers to the area enclosed by the boundary of the V-I trajectory. It is proportional to the magnitude of phase angle difference between voltage and current waveforms. Area enclosed by line segment *a* in figure 5, represents this metric.

**Non-linearity of Mean Curve:** Mean curve of the V-I trajectory, represented in fig. 5 by line segment *b*, divides it into two identical halves. Degree of distortion of the mean-curve from a straight line is indicative of the non-linearity of electrical behavior of a particular appliance load. An estimate of this metric is the area enclosed under mean curve *b* when viewed with reference to line segment c.

**Number of Self-Intersections:** Number of self-intersections for a V-I trajectory also corresponds to presence of higher order harmonics in current waveform.

**Slope of Middle Segment:** A near-zero slope of middle segment of the mean curve *b* is often characteristic of power-electronic loads and serves to differentiate them from other kinds of loads. In figure 5, tangent of the angle between line segments *b* and V-axis ($\theta$) represents this metric.

**Area of Right and Left Segments:** This refers to area enclosed by near-vertical segments towards the edges of the V-I trajectory.

We also propose a new shape feature of the V-I trajectory we refer to as *span* of the V-I trajectory. It represents the vertical distance between highest and lowest I-coordinate, equal to length of vertical component of line segment c in fig. 5. It has a correspondence with fundamental active power *P* as it increases proportional to amplitude of instantaneous current.

TABLE I
LINEAR CORRELATION COEFFICIENTS - WS AND PQ

| LS | areaRL, $P_{rms}$ | curveML, TeHD | numIntersec, ToHD | numIntersec, areaEnc | span, $P_{rms}$ |
|---|---|---|---|---|---|
| Coeff. | 0.5096 | 0.6038 | 0.6861 | -0.5215 | 0.96 |

Using this shape difference we are able to differentiate between switching events from various appliances. Table I expresses linear correlation coefficients for some of the aforementioned WS and PQ quantities. A magnitude close to one expresses strong linear dependence between two variables while a magnitude close to zero implies minimal linear relationship between the two. Coefficients in table I have been estimated using unique combinations from a set of about fifty thousand switching events belonging to REDD. A strong linear proportionality between curvature of mean line and harmonic content in current waveform expressed by $TeHD$ is particularly obvious from the table. An increase in area enclosed by V-I trajectory is typically accompanied by a decrease in the number of self-intersections, illustrated by a negative value for respective coefficient in table I.

### D. Disaggregation Algorithms

Choice of learning algorithm is a function of the characterization of the learning problem and the load signature used. The inputs to each learning algorithm are a specific number of training examples; each example is a feature vector representing particular load metrics coupled with respective appliance label, ($\mathbf{x_i}$, y), $\mathbf{x_i} \in \mathbb{R}^F$. For instance, HAR training example for each switching event contains 77 real numbers representing spectral power for concurrent segments of frequency for FT of current waveform alongside an appliance label. These 77 real numbers are extracted from FT for a range of 0-4kHz. WS contains 7 real numbers, PQ contains 4 real numbers. Feature vectors corresponding to switching events are divided into three sets namely training, cross-validation and, test set. Training and cross-validation examples are used by learning algorithm to produce a generalized model subsequently used to predict labels for test set feature vectors.

In our evaluation we have used four learning algorithms to evaluate our approach. These four algorithms represent popular choices of off-the-shelf learning algorithms proven effective for a wide variety of classification problems across domains ([21], [33]).

First of these algorithms is a feed-forward artificial neural network (ANN) trained using Levenberg-Marquardt (LM) method [18]. A detailed discussion on ANN topologies can be found in [18].

The second algorithm is a hybrid learning algorithm (ANN + EA), composed of an artificial neural network coupled with an evolutionary algorithm (EA with momentum) block for local search. Once the weights and parameters for the ANN part are determined by LM or any other training algorithm [18], the EA part conducts a local search around the final solution for better prediction accuracy. The principle for evolution from one generation of parameters $W_n$ to the next, $W_{n+1}$, is expressed by the following equations [21].

$$\Delta W_{n+1} = m \, \Delta W_n + (1-m) g W_n \quad (3)$$

$$W_{n+1} = W_n + \Delta W_{n+1} \quad (4)$$

In 3 and 4, momentum constant $m$ is typically a real number from the range $(0,1)$ that describes the degree or threshold of 'presence' of previous generation of parameters $W_n$ in the next generation $W_{n+1}$. Method to ascertain value of momentum constant $m$ using stochastic global search is described in the next subsection. The value of $g$, a constant real number, is selected from the range $(0, 0.1)$ for all generations of EA [21].

The third algorithm evaluated for the course of this study is support vector method (SVM) with a Gaussian kernel function. Intuitively, the kernel function describes a distance measure that suggests weightage for various training instances in order to construct a hypothesis that determines the class label for each training instance. SVM theory and application to MS-NILM for a relatively small set of appliance feature instances is discussed in more relative detail by Onoda et. al. [31].

The fourth and final algorithm is Adaptive Boost (AdaBoost) that uses decision stumps as weak classifiers. The premise of AdaBoost is to iteratively boost the prediction performance of the weak classifier over the training instances. This is accomplished by maintaining a distribution of weights that is updated in successive iterations so as to 'concentrate' on instances that are misclassified in the previous iterations [32].

Liang et. al. [8] and Onoda et. al. [31] have demonstrated the use of ANN, SVM and Adaptive Boost respectively in context of NILM under separate system configurations. No systematic examination of all four with WS over benchmark datasets exists so far, hence these have been employed for this study.

*E. Model Selection and Performance Optimization*

Prior to training of all the aforementioned learning algorithms, setup parameters (number of neurons in each hidden layer $N_h$ in case of ANN, momentum constant $m$ in case of EA) have to be specified. This step is typically referred to as 'model selection' ([23], [24]). In the course of this study, an enhanced variant of differential evolution (EDE) is used to automatically search and select these setup parameters. Figure 7 illustrates the model selection step in detail. A specific number of randomly chosen setup parameters or a candidate 'population' of parameters is chosen to be evaluated for prediction accuracy. The population is altered or 'mutated' and a new population is selected by selectively combining the initial and altered population according to specific rules (describing the 'fitness' of altered population relative to that of original population) so as to gradually maximize the prediction accuracy or minimize prediction error in the new population as this cycle is repeated. The term 'objective function' refers to prediction error in the following passage.

Differential Evolution (DE) is a heuristic, population-based global search strategy that offers more relative certainty and efficiency of convergence for minimization problem for non-linear continuous space functions [6]. Following passage reviews the algorithm used for parameter search ([4], [12]),

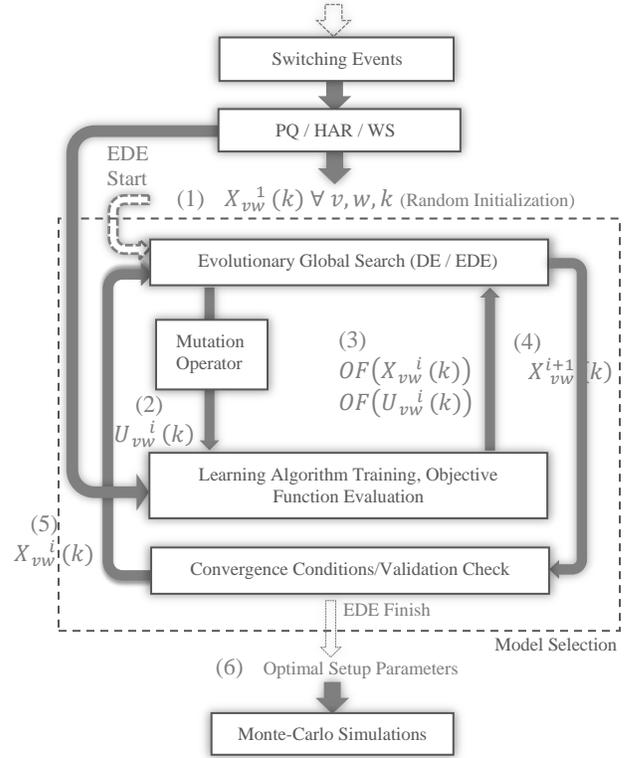

Fig. 6. Global Search using enhanced differential evolution for model selection in ANN, ANN + EA and SVM.

with special reference to DE and EDE.

DE prescribes an empirical constant RR (recombination rate) that decides the threshold for combining altered and original populations of parameters. Choice of this constant is a function of observer's experience and expertise. EDE is an enhanced variant of DE proposed in [4] whereby instead of an empirical recombination rate (RR), a 'fitness function' is described that weighs the fitness of mutant population relative to fitness of original population,

$$\text{Fitness of Mutant Population} = \frac{\frac{1}{OF(U)}}{\frac{1}{OF(X)} + \frac{1}{OF(U)}} \quad (5)$$

$$\text{Fitness of Original Population} = \frac{\frac{1}{OF(X)}}{\frac{1}{OF(X)} + \frac{1}{OF(U)}} \quad (6)$$

In eqs. 5 and 6, 'U' represents the mutant population of training parameters, 'X', the original population, 'OF', the objective function corresponding to a set of ANN training parameters. So, a constant RR is replaced by a dynamic RR that weighs how much the altered population excels the original population in prediction accuracy.

A brief summary of the overall proposed strategy is described as follows. Fig. 6 illustrates the sequence of steps involved in the EDE-based global search algorithm, labeled as follows.

1) Populations of system variables (number of hidden layer neurons in case of ANN, recombination rate in case of DE/GA or momentum constant in EA) or

'genes' for $k$th trainer and $i$th iteration, $X_{vw}^i(k)$, $k = 1, 2, \ldots, N$ are randomly initialized in the beginning, a total of M individuals ($v = 1, 2, \ldots, M$) and G genes per individual for each trainer ($w = 1, 2, \ldots, G$).

2) Each individual in the mutant population $U_{vw}^i(k)$ is determined by a linear combination of genes from three randomly chosen individuals in the original population.

$$U_{vw}^i(k) = X_{aw}^i(k) + F \times \left(X_{bw}^i(k) - X_{cw}^i(k)\right)$$
$$a \neq b \neq c \neq v, \quad F > 0$$

3) Objective function values for mutant and original population are determined by evaluating prediction error for the multi-class classifier in question (for a specific feature space and a specific learning algorithm). Fitness functions of mutant and original populations are determined from 1 and 2.

4) Each gene in $X_{vw}^{i+1}(k)$ is determined as follows:

$$x_{vw}^{i+1}(k) = \begin{cases} u_{vw}^i(k) & \text{if } rand < Fitness\ Function\ (U) \\ x_{vw}^i(k) & \text{if } rand \geq Fitness\ Function\ (U) \end{cases}$$

5) Step 2 is repeated with $X_{vw}^{i+1}(k)$ in place of $X_{vw}^i(k)$ until one of the stopping conditions is met, that is, either the objective function is minimized beyond a specific threshold (a real number) or it stays constant for a specific number of consecutive iterations (resulting in a 'validation stop') or the maximum number of iterations is reached.

6) After one of the stopping conditions is met, the best out of M individuals in the population $X_{vw}^{i+1}(k)$ – the one with the least value for objective function – is chosen to be the optimal set of setup parameters for learning algorithm in question.

For a multi-class classifier, one choice is that performance optimization of each class can be carried out in parallelized fashion (requiring $N$ replicas of above algorithm) such that for selected values of $k$,

$$\left|OF_k - \frac{1}{N}\right| < \varepsilon$$

An alternative choice would be to use a single objective function following the above rule; simplest instances would be for selected values of $k$, $\sum_k \left|OF_k - \frac{1}{N}\right|$ or $\sum_k (OF_k - \frac{1}{N})^2$, depending on the penalty required as error grows and whether $OF$ can adapt to degree of convergence of parameters like with the basic case of $\max_k OF_k$. However, design of sophisticated objective functions that exhibit both properties (dynamic adaptation and selective minimization) is outside the scope of this study and will be addressed in subsequent research.

## IV. Experimental Evaluation

This section presents numerical evaluation of benchmark signatures and learning algorithms for precision of prediction of appliance labels for switching events extracted from REDD.

We have evaluated load signatures corresponding to these switching events for WS, PQ, HAR using the algorithms reviewed earlier: artificial feed-forward neural network (ANN), hybrid neural network (ANN + EA), support vector machine (SVM) and adaptive boost (AdaBoost) for one-dimensional decision stumps.

Precision of prediction, $\eta \in [0, 1]$, used as accuracy metric, is determined as the number of correctly predicted classes, $P_c$, for extracted switching events in the database, weighted by the total number of switching events, $P_t$. This study assumes perfect recall, that is, all true switching events are detected and reported to learning algorithms.

$$\eta = \frac{P_c}{P_t} \quad (7)$$

### A. Description of Dataset and Initial Conditions

Reference Energy Disaggregation Dataset (REDD) is a publicly available dataset containing detailed energy usage information of several homes over extended periods and in two granularities [14]. The low granularity data is average real power consumption of multiple households (both the mains and individual circuits) at a frequency of approximately 1Hz for mains and 0.33Hz for individual circuits. High granularity data is AC voltage and current waveform data from household mains acquired using commercial load monitors at a frequency of 16.5 kHz.

REDD high frequency data considered for this evaluation contains energy usage data worth about twenty days for two houses. In case of house # 3, for instance, a total of 22 channels of voltage and current waveform data are present. Prior to clustering the delta-form signatures, only appliances with $P > 50W$ are considered.

TABLE II
CIRCUIT LABELS FOR REDD HOUSEHOLD # 3 – 22 CHANNELS

| Cct. Label | Appliance | Cct. Label | Appliance | Cct. Label | Appliance |
|---|---|---|---|---|---|
| 1, 2 | mains | 9 | dishwasher | 16 | microwave |
| 3 | unknown | 10 | furnace | 17 | lighting |
| 4 | unknown | 11 | lighting | 18 | Smoke alarms |
| 5 | lighting | 12 | unknown | 19 | lighting |
| 6 | electronics | 13 | Washer-dryer | 20 | Bathroom gfi |
| 7 | refrigerator | 14 | Washer-dryer | 21 | Kitchen outlets |
| 8 | disposal | 15 | lighting | 22 | Kitchen outlets |

Subsequent to signature extraction, relative sizes of 0.45, 0.1 and 0.45 are utilized for training, cross-validation and test sets in ANN training, a base case for further comparison in next section. An early stopping condition is in effect to avoid over-fitting and keep the ANN, ANN+EA predictors generalized. Over-fitting refers to near-perfect precision for training set examples ($\eta \approx 1$) while poor precision for test set examples – generally a characteristic of a high variance for learning algorithm weights and biases. For EDE, F = 0.5 is generally a good choice with both ANN and ANN+EA. EDE (typically 30 individuals, 50 iterations) is utilized to assert the best possible training parameters. Subsequently Monte Carlo simulations are conducted with these parameters to evaluate $\eta$.

*B. Monte-Carlo Simulations*

A Monte-Carlo simulation evaluates the precision of employed learning algorithms a large number of times under an unaltered computational capacity for the duration of the simulation. In case of ANN as learning algorithm for instance, for every iteration of the simulation, Levenberg-Marquardt (LM) algorithm is invoked to search for ANN weights (using the number of neurons in each hidden layer from the previous model-selection step). The large number of simulations accounts for variability of precision of prediction due to random initialization of weights during parameter search for learning algorithms.

Available degrees of freedom while pre-processing the datasets in order to extract switching events are:
- Relative sizes of training and test sets,
- threshold on active appliance power consumption ($P_{min}$) and purity of clustering done during pre-processing [19]

Unique combinations of these criteria can be modified to generate a unique combination of switching events from the dataset and a unique Monte-Carlo simulation.

Relative sizes of training and test sets compare the robustness of performance for various algorithms over PQ, HAR and WS, particularly during early operational periods in an event-based NILM system where training examples are too few for the learning algorithm to account for all possible load scenarios and consequently to yield generalizable appliance models.

Purity of clustering [19] during the pre-processing module is related to the optimal number of appliance classes selected in order to cluster the extracted waveforms. This optimal number of classes is a function of threshold on power consumption of appliances, $P_{min}$, that are included in a particular Monte-Carlo simulation. A large number of appliances with values of load signatures in close proximity (high 'similarity'), noisy and/or highly non-sinusoidal conditions might lead to a larger number of anomalous training examples ('outliers') that might not represent the most frequent operating state for the appliance in question, which might render the learning algorithm more susceptible to incorrect classification, as will be demonstrated in the next section.

It is important to indicate here that these two function as so-called *extrinsic* criteria for generating unique load scenarios, in the sense that these are modifiable while retaining the original sequence of occurrence of switching events and do not represent intrinsic load characteristics. In section V, this original sequence of switching events from REDD is modified. A large number of unique load scenarios are generated using appliance profiles (feature vectors) from REDD. This allows us to simulate the effects of dynamic and noisy load signatures on capacity of classification for various algorithms and load signatures. Research by Liang and colleagues [8] presents convenient criteria for load dynamics, appliance similarity and electrical noise that are consulted for the purpose of this study. Definitions for these criteria *intrinsic* to composite load are mentioned in the next section.

*C. Numerical Results*

Table III lists the median overall, training set and test-set prediction accuracy for the three benchmark load signatures (an Intel Core i7 machine, CPU clock 3.1 GHz, 8GB of RAM). It is evident from the numerical figures in table III that wave-shape metrics (WS) outperform or generally compare with both PQ and HAR in prediction accuracy. It is important to indicate here that since HAR is represented by 77 real numbers expressing spectral power for various segments of the Fourier transform while WS is represented by 7 real numbers for metrics described in the previous section so for orders of magnitude less number of features, WS outperforms or at a minimum, offers comparable prediction accuracy with HAR and hence is relatively more robust for event-based operation. This is also consistent with the fact that for more than 90% of all training examples used in these simulations, above 80% of spectral energy in HAR appears in segments 0 Hz – 600 Hz and 3 kHz – 4.5 kHz and intermediary segments do not add meaningful information to individual feature vectors so the 'effective' dimensions of HAR are less than 77 and correspondence between WS and HAR (presence of higher order harmonics indicated by self-intersections of V-I trajectory etc.) is captured in the aforementioned frequency bands.

It is also evident that including EA with momentum alongside ANN as learning algorithm does not provide a substantial improvement in performance over ANN.

TABLE III
MEDIAN OVERALL/TEST SET/TRAINING SET PREDICTION ACCURACY (%) FOR VARIOUS CHOICES OF LOAD SIGNATURE FROM MONTE-CARLO SIMULATIONS

| LS | ANN | ANN + EA | SVM | AdaBoost |
|---|---|---|---|---|
| PQ | 88.5/88.3/88.8 | 88.8/88.6/88.9 | 97.2/96.9/97.5 | 99.3/98.8/98.8 |
| HAR | 82.1/80.4/83.4 | 82.8/81.4/84.4 | 98.7/98.0/99.3 | 98.8/97.4/99.8 |
| WS | 92.0/91.5/92.5 | 91.5/90.9/92.1 | 98.1/97.1/98.9 | 99.1/98.7/99.1 |

Table IV lists the maximum, mean and median precision for ANN, ANN+EA, SVM and AdaBoost over PQ, HAR and WS for respective Monte-Carlo simulation.

TABLE IV
OVERALL MAXIMUM/MEAN/MEDIAN PREDICTION ACCURACY (%) FOR VARIOUS CHOICES OF LOAD SIGNATURE FROM MONTE-CARLO SIMULATIONS

| LS | ANN | ANN + EA | SVM | AdaBoost |
|---|---|---|---|---|
| PQ | 91.1/88.4/88.5 | 91.7/88.9/88.8 | 97.4/97.3/97.2 | 99.4/99.3/99.3 |
| HAR | 87.1/81.1/82.1 | 88.0/82.3/82.8 | 98.8/98.7/98.7 | 98.9/98.8/98.8 |
| WS | 94.3/91.4/92.0 | 96.2/90.8/91.5 | 98.3/98.1/98.1 | 99.2/99.1/99.1 |

Fig. 7 indicates an almost comparable resilience to increased proportion of test set data for PQ, HAR and WS with AdaBoost and ANN. Fig. 8 indicates the drop in precision of prediction with smaller values of $P_{min}$, for ANN as learning algorithm. This generally corresponds to a large number of low-power load signatures in close proximity, as discussed previously.

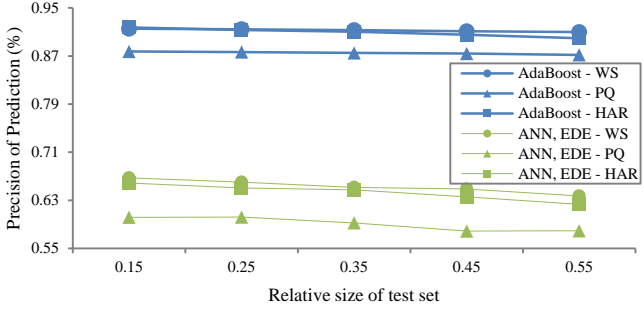

Fig. 7. Trends of precision of prediction for learning algorithms as a function of relative size of training, test and cross-validation sets.

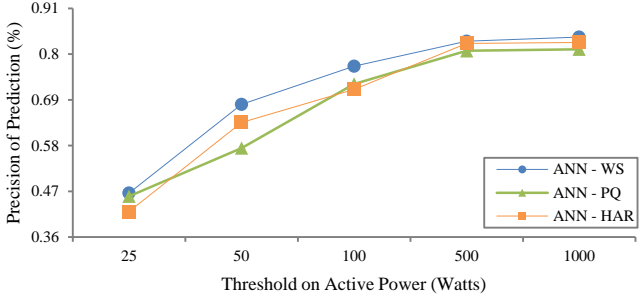

Fig. 8. Trends of precision of prediction for learning algorithms as a function of minimum threshold on appliance power consumption.

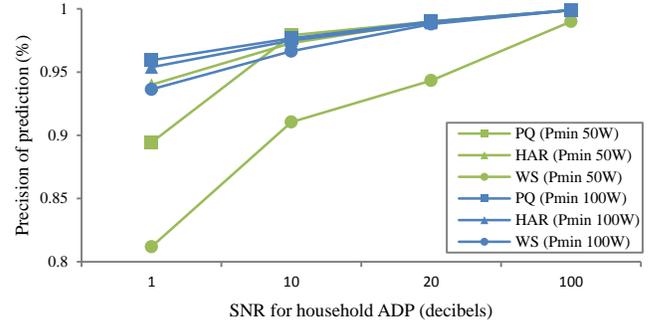

Fig. 9. Precision of prediction as a function of signal-to-noise ratio for household ADP using Adaptive Boost as classification algorithm.

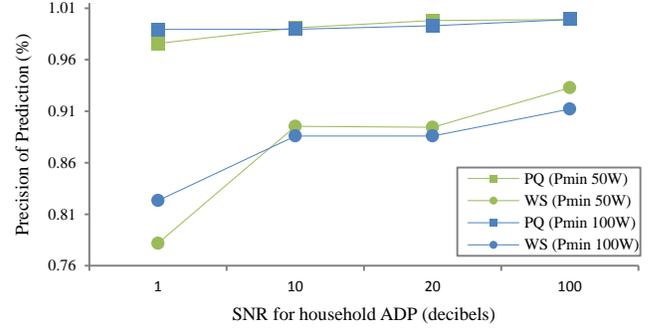

Fig. 10. Precision of prediction as a function of signal-to-noise ratio for household ADP using Support Vector Method as classification algorithm.

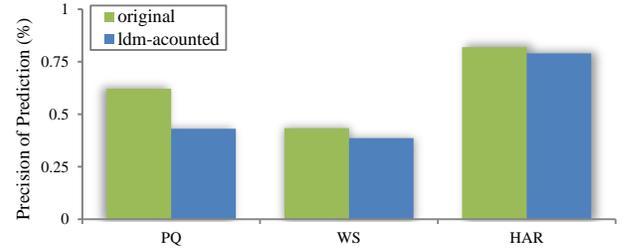

Fig. 11. Disparity in precision of prediction for various load signatures and classification algorithms with inclusion of load dynamics

## V. EXTENDED SIMULATIONS

In the next portion of our empirical study, appliance load profiles (feature vectors) derived from REDD are used to generate simulations for thousands of different load scenarios. Each load scenario contains a sequence of switching events for a defined period of time, typically ranging from a day to a month. These scenarios are generated by sequential addition/subtraction of individual signatures to/from the composite load. The signatures from our database contain as many as 40 distinct appliance signatures for different respective operating states.

The simulation accounts for dynamic loading of various ubiquitous loads (particularly household electric motors and air conditioners) and electrical noise. A complete description of implemented NILM simulator is outside the scope of this paper and will be detailed in a future study. However, the relevant results of experiments pertaining to the degrees of freedom affecting the precision of prediction for various algorithms with WS are detailed.

### A. Sensitivity towards Noise in Load Signatures

Effect of noise is simulated by adding zero mean white Gaussian noise to ADP of generated load scenarios. For this particular set of Monte-Carlo simulations ($P_{min}$ = 50W, 100W, 14 days, 15 switching events per hour with a normally distributed frequency), all employed algorithms classify more than 90% of training instances of load signatures for SNR > = 20. However, precision of prediction drops with decrease in SNR, most considerably for WS which suffers more drastic drop in precision compared to PQ and HAR for all employed algorithms. Figs. 9 and 10 illustrate the case for Adaptive Boost and Support Vector Method. About 20% drop in case of adaptive boost (with $P_{min}$ = 50W which corresponds to about 12000 training instances) is evident from fig. 10 (a). This type of response is associated with drastic shape distortion of V-I trajectories of respective appliances for small values of SNR.

### B. Sensitivity towards Dynamic Load Signatures

Dynamic behavior of appliance loads is recognized as inherent variability of shape for instantaneous current waveform due to of dynamic loading. Similarity refers to the proximity of numerical values of load signatures. Discrete Fourier transform ($X(f_k) = k_r + jk_i$) of various single-cycle snapshots (CW) of an appliance load is calculated using FFT algorithm. Range of values for $k_r$ and $k_i$ in the database expresses the degree of variability in CW for respective appliances. Inverse Fourier transform is used to reconstruct CW for appliance signatures keeping in account the updated $k_r$ and $k_i$ values weighted by LDM. A modification of the relationship between reconstructed and original CW proposed by Liang et. al. [8] is employed for this study.

$$CW_i(t) = \frac{1}{2}\sum_{c=0}^{N-1}[(sel(k_{r,c})) + j(sel(k_{i,c}))]e^{\frac{j2\pi ct}{N}} \quad (8)$$

In eq. 8, $sel(k_{r,c})$ and $sel(k_{i,c})$ refer to a uniform sampling function that randomly selects any one of the available $k_r$ and $k_i$ in the database for a particular appliance, in order to produce all possible variations of the instantaneous current waveform with a uniform probability in the updated CW used for subsequent Monte-Carlo simulations. Fig. 11 highlights the disparity in precision of prediction for original and adjusted CW, AdaBoost and PQ, WS and HAR over a large number of simulated load scenarios. PQ generally suffers a slightly larger disparity in precision compared to HAR and WS for this set of Monte-Carlo simulations ($P_{min}$ = 25W, 20 days, SNR = 10, 15 switching events per hour with a normally distributed frequency). A larger separation between signature clusters in case of WS and HAR might be associated with this particular observation, so as to allow for a relatively robust precision of prediction. However, this difference in precision for WS and PQ diminishes for smaller values of SNR where WS suffers an explicit disadvantage obvious from the previous subsection.

## VI. CONCLUSIONS AND FUTURE WORK

As was demonstrated in earlier sections, wave-shape metrics (WS) offer superior or generally comparable performance in load disaggregation under tested scenarios and have a direct correspondence to operating characteristics of appliances as contained in current wave-shape. This establishes a promising direction for research in unsupervised energy disaggregation and fault monitoring of devices and appliances with WS in real-life residential and commercial settings. Model selection and performance optimization for single-algorithm and ensemble predictors designed for MS-NILM and AS-NILM is an associated promising direction for subsequent research. Applications of metaheuristic optimization and online learning in NILM can prove instrumental in design of state-of-the-art analytic applications that leverage modern service architectures for increased productivity, reliability and scalability envisioned for electric grid of the future.

## REFERENCES

[1] G.W. Hart, "Non-intrusive load monitoring," *Proceedings of the IEEE*, vol. 80, no. 12, pp. 1870-1891, Dec 1992.
[2] J. Liang, S.K.K. Ng, G. Kendall and J.W.M. Cheng, "Load signature study part 1: basic concept, structure and methodology," *IEEE Trans. Power Delivery,* vol. 25, no. 2, pp. 551-560, April 2010.
[3] H.Y. Lam, G.S.K. Fung, and W.K. Lee, "A Novel Method to Construct Taxonomy of Electrical Appliances Based on Load Signatures," *IEEE Trans. Consumer Elec.,* vol.53, no.2, pp.653-660, May 2007.
[4] N. Amjady, F. Keynia and H. Zareipour, "Short-term load forecast of microgrids by a new bilevel prediction strategy," *IEEE Trans. Smart Grid*, vol.1, no.3, pp.286-294, Dec. 2010.
[5] S. Shaw, S. Leeb, L.K. Norford and R. Cox, "Nonintrusive load monitoring and diagnostics in power systems," *IEEE Trans. Instrument. & Measurement,* vol.57, no.7, July 2008.
[6] R. Storn and K. Price, "Differential Evolution - A simple and efficient heuristic for global optimization over continuous spaces," *Journal of Glob. Optimization,* vol. 11, no. 4, pp. 341-359, Dec. 1997.
[7] A. I. Cole and A. Albicki, "Non-intrusive Identification of Electrical Loads in a Three-Phase Environment based on Harmonic Content," *Proc. IEEE Instrumentation and Measurement Technology Conference*, pp. 24-29, 2000.
[8] J. Liang, S.K.K. Ng, G. Kendall and J.W.M. Cheng, "Load Signature Study - Part II: Disaggregation Framework, Simulation and Applications," *IEEE Trans., Power Delivery,* vol.25, no.2, pp.561-569, April 2010.
[9] H. W. Lai, G.S.K. Fung, H.Y. Lam and W.K. Lee, "Disaggregate Loads by Particle Swarm Optimization Method for Non-intrusive Load Monitoring," *International Conference on Electrical Engineering,* July 2007.
[10] K. Suzuki, S. Inagaki, T. Suzuki, H. Nakamura, and K. Ito, "Nonintrusive appliance load monitoring based on integer programming," *SICE Annual Conference*, pp.2742-2747, Aug. 2008.
[11] Y. H. Lin, M. S. Tsai, and C. S. Chen. "Applications of fuzzy classification with fuzzy c-means clustering and optimization strategies for load identification in NILM systems," *Proc. IEEE International Conference on Fuzzy Systems*, pp. 859-866, 2011.
[12] T. Hassan, "Bi-level characterization of manual setup non-intrusive demand disaggregation using enhanced differential evolution," presented at the 1at Int. Workshop on Non-Intrusive Load Monitoring, Pittsburgh, PA, USA, 2012.
[13] A. Qing, *Differential Evolution, Fundamentals and Applications in Electrical Engineering*, Wiley, 2009.
[14] J. Zico Kolter and Matthew J. Johnson, "REDD, A public dataset for energy disaggregation research," *Proceedings of the SustKDD workshop on Data Mining Applications in Sustainability*, 2011.
[15] M. Bergs, E. Goldman, H. Scott Matthews and L. Soibelman. "Enhancing Electricity Audits in Residential Buildings with Non-Intrusive Load Monitoring," *Journal of Industrial Ecology: Special Issue on Environmental Applications of Information and Communication Technologies*, vol. 14, no. 5, pp. 844-858, 2010.
[16] K.D. Lee, "Electric Load Information System based on Non-Intrusive Power Monitoring," Ph.D. dissertation, Dept. Mech. Eng., Massachusetts Institute of Technology, 2003.
[17] S. Leeb, "A Conjoint Pattern Recognition Approach to Non-Intrusive Load Monitoring," Ph.D. dissertation, Dept. Elec. Eng.& Comp. Sci., Massachusetts Institute of Technology, 1993.
[18] M. Beale and H. Demuth, "Neural Networks Toolbox for Use with MATLAB: User's Guide," Ver. 3, Massachusetts: The MathWorks, 1998.
[19] Christopher D. Manning, Prabhakar Raghavan and Hinrich Schütze, "Introduction to Information Retrieval," Cambridge University Press. 2008.
[20] IEEE Std. 1459, "IEEE Standard Definitions for Measurement of Electric Power Quantities under Sinusoidal, Non-Sinusoidal, Balanced or Unbalanced Conditions", February 2010.
[21] N. Amjady and F. Keynia, "Day-Ahead Price Forecasting of Electricity Markets by Mutual Information Technique and Cascaded Neuro-Evolutionary Algorithm," *IEEE Trans. Power Systems,* vol. 24, no. 1, pp. 306-318, February 2009.
[22] LK Norford, SB Leeb, "Non-intrusive Electrical Load Monitoring in Commercial Buildings based on steady state and transient load detection algorithms," *Energy and Buildings*, 1996.
[23] U. Anders and O. Korn. "Model selection in neural networks," *Neural Networks*, vol. 12, no. 2, pp. 309-323, 1999.
[24] H.K.H. Lee, "Model selection for neural network classification," *Journal of Classification*, vol. 18, no. 2, pp. 227-243, 2001.
[25] O. Parson, S. Ghosh, M. Weal and A. Rogers, "Non-Intrusive Load Monitoring using Prior Models of General Appliance Types," *Proc. Twenty-Sixth Conf. Artificial Intelligence*, pp. 356-362, 2012.
[26] H. S. Kim, "Unsupervised Disaggregation of Low Frequency Power Measurements," M.S. thesis, Dept. Comp. Sci., University of Illinois Urbana-Champaign, 2012.
[27] Z. Ahmed, A. Gluhak, M. A. Imran, and S. Rajasegarar, "Non-Intrusive Load Monitoring Approaches for Disaggregated Energy Sensing: A Survey," *Sensors*, vol. 12, no. 12, pp. 16838-16866, 2012.
[28] M. Zeifman and K. Roth. "Nonintrusive appliance load monitoring: Review and outlook," *IEEE Trans. Consumer Electronics*, vol. 57, no. 1, pp. 76-84, 2011.
[29] C. Laughman, K. Lee, R. Cox, S. Shaw, S. Leeb, L. Norford, and P. Armstrong. "Power signature analysis," *IEEE Power and Energy Magazine*, vol. 1, no. 2, pp. 56-63, 2003.

[30] K. D. Lee, , S. B. Leeb, L. K. Norford, P. R. Armstrong, J. Holloway, and S. R. Shaw. "Estimation of variable-speed-drive power consumption from harmonic content," *IEEE Trans. Energy Conversion*, vol. 20, no. 3, pp. 566-574, 2005.

[31] T. Onoda, G. Rätsch, and K. R. Müller. "Applying support vector machines and boosting to a non-intrusive monitoring system for household electric appliances with inverters," *Second Int. ICSC Symposium on Neural Computation*, 2000.

[32] Y. Freund, R. Schapire, and N. Abe. "A short introduction to boosting," *Journal Japanese Soc. Artificial Intelligence*, vol. 14, pp. 771-780, 1999.

[33] I. Guyon. (2013, January 10). SVM Application List [Online]. Available: http://www.clopinet.com/isabelle/Projects/SVM/applist.html

[34] C. W. Gellings, "The concept of demand-side management for electric utilities," *Proceedings of the IEEE*, vol. 73, no. 10, pp. 1468-1470, 1985.